\documentclass[11pt,letterpaper]{article}
\usepackage[utf8x]{inputenc}
\usepackage[T1]{fontenc}
\usepackage{mathptmx} 
\usepackage{authblk}
\usepackage[pdftex]{graphicx} 
\usepackage[pdftex,linkcolor=black,pdfborder={0 0 0}]{hyperref} 
\usepackage{calc} 
\usepackage{enumitem} 
\usepackage{xcolor}
\usepackage{enumitem}
\usepackage{xurl}
\usepackage{booktabs}
\usepackage{float} 
\usepackage[numbers]{natbib}
\usepackage[all]{nowidow} 
\usepackage[protrusion=true,expansion=true]{microtype} 
\usepackage[letterpaper, lmargin=0.1\paperwidth, rmargin=0.1\paperwidth, tmargin=0.1\paperheight, bmargin=0.1\paperheight]{geometry} 
\usepackage[font=small, labelfont=bf, aboveskip=5pt]{caption} 
\usepackage{makecell}
\usepackage{amsmath}

\newcommand{\YCjmt}{Y_{jmt}(0)}
\newcommand{\YhatCimt}{\hat{Y}_{imt}(0)}
\newcommand{\YhatCimtBC}{\hat{Y}^{BC}_{imt}(0)}
\newcommand{\YhatCdotmtBC}{\hat{Y}^{BC}_{\cdot mt}(0)}
\newcommand{\YTdotmt}{Y_{\cdot mt}(1)}
\newcommand{\Mmt}{\hat{\mu}^{(0)}_{mt}}
\DeclareUnicodeCharacter{2212}{-}

\begin{document}

\title{Community notes reduce engagement with and\\diffusion of false information online}

\author[a]{Isaac Slaughter}
\author[b]{Axel Peytavin}
\author[b,c]{Johan Ugander}
\author[a,1]{Martin Saveski}

\renewcommand\Affilfont{\small} 
\renewcommand\Authfont{\large}

\affil[a]{Information School, University of Washington}
\affil[b]{Department of Management Science and Engineering, Stanford University}
\affil[c]{Department of Statistics and Data Science, Yale University}
\affil[1]{Corresponding author: \url{msaveski@uw.edu}}

\date{\vspace{−5ex}}

\maketitle

\begin{abstract}
Social networks scaffold the diffusion of information on social media. Much attention has been given to the spread of true vs.\ false content on online social platforms, including the structural differences between their diffusion patterns. However, much less is known about how platform interventions on false content alter the engagement with and diffusion of such content. In this work, we estimate the causal effects of Community Notes, a novel fact-checking feature adopted by X (formerly Twitter) to solicit and vet crowd-sourced fact-checking notes for false content. We gather detailed time series data for 40,078 posts for which notes have been proposed and use synthetic control methods to estimate a range of counterfactual outcomes. We find that attaching fact-checking notes significantly reduces the engagement with and diffusion of false content. We estimate that, on average, the notes resulted in reductions of 46.1\% in reposts, 44.1\% in likes, 21.9\% in replies, and 13.5\% in views after being attached. Over the posts' entire lifespans, these reductions amount to 11.6\% fewer reposts, 13.3\% fewer likes, 6.9\% fewer replies, and 5.5\% fewer views on average. In reducing reposts, we observe that diffusion cascades for fact-checked content are less deep and less ``viral,'' but not less broad, than synthetic control estimates for non-fact-checked content with similar reach. This structural difference contrasts notably with differences between false vs.\ true content diffusion itself, where false information diffuses farther, but with structural patterns that are otherwise indistinguishable from those of true information, conditional on reach.\footnote{
Accepted for publication in the \textit{Proceedings of the National Academy of Sciences}. 
Changes from prior \textit{arXiv} version:
title updated in response to reviewer comment; 
added robustness checks, permutation test, and expanded discussion; 
corrected software issue, leading to changes in magnitude of unconditional average effects by no more than 8\% (no changes to statistical significance or direction).
Published version DOI: \url{https://doi.org/10.1073/pnas.2503413122}
}
\end{abstract}

\section*{Introduction}
The spread of false information on social media poses risks to public health~\cite{pennycook2020fighting}, democratic processes~\cite{ecker_misinformation_2024}, and social cohesion~\cite{garrett2019partisan}. Social media has been broadly observed to preferentially support the spread of false news over true news~\cite{friggeri_rumor_2014,vosoughi2018spread,juul2021comparing,gonzalez2024diffusion}. Scholars as well as social media platforms are actively working to design and test strategies to limit its transmission \cite{lazer2018science,bak2022combining,kozyreva2024toolbox}, including fact-check warning labels placed on individual sources or pieces of information~\cite{tameez2020,aslett2022news}, educational interventions to boost users’ competencies at identifying false information~\cite{roozenbeek2019fake,guess2020digital, sirlin2021digital, moore2022digital}, and a shift to design objectives other than user engagement~\cite{milli2021optimizing,bernstein2023embedding,cunningham2024we,piccardi2024social}. 

Professional fact-checking is the most widely used intervention against misinformation, often implemented by attaching warning labels to fact-checked posts~\cite{mosseri2016news, instagram2019combatting,twitter2019yoel}. Studies investigating the effectiveness of these labels find that they decrease self-reported belief in and willingness to share misinformation~\cite{martel2024fact, porter2022political, clayton2020real}. However, even if effective, professional fact-checking is costly and difficult to scale both in speed and coverage \cite{wack_political_2024}, and increasingly viewed with skepticism by segments of the public~\cite{pew2020factchecking}. Crowd-sourced fact-checking has emerged as a promising alternative, leveraging the ``wisdom of the crowd,’’ i.e., that aggregating judgments of groups of non-experts leads to accurate assessments even if the individual assessments are inaccurate~\cite{galton1949vox}. Lab experiments investigating the feasibility of crowd-sourced fact-checking find that groups as small as 15 people can identify misinformation as accurately as professional fact-checkers~\cite{allen2021scaling, bhuiyan2020investigating, resnick2023searching, arechar2023understanding}.

Building on these findings, X (formerly Twitter) introduced a crowd-sourced fact-checking system called Community Notes~\cite{coleman2021introducing}. The system enables ordinary users to propose fact-checking notes to be attached to potentially misleading posts and rate the helpfulness of proposed notes. The system uses a ``bridging-based'' matrix factorization algorithm to score the overall helpfulness of notes based on the individual ratings~\cite{wojcik_birdwatch_2022}. Notes rated helpful by many users with diverse views, as measured by estimated latent positions, are scored higher. Only notes that cross a certain helpfulness threshold are classified helpful and displayed with the post.

Upon introducing the Community Notes program, X reported results from an A/B test that notes selected by the bridging-based algorithm reduced individual-level decisions to like and repost misinformation by 25-−34\% relative to a control group~\cite{wojcik_birdwatch_2022}. Through the lens of widely employed epidemiological models of information diffusion, changes in the probability that individual units will share content typically have a highly non-linear relationship with the overall number of people exposed to the content~\cite{larremore2021test}, a quantity that is not easily assessed through an A/B test due to ubiquitous network effects, i.e., interference between treatment and control units~\cite{eckles2017design}.

\begin{figure*}
\centering
\includegraphics[width=\linewidth]{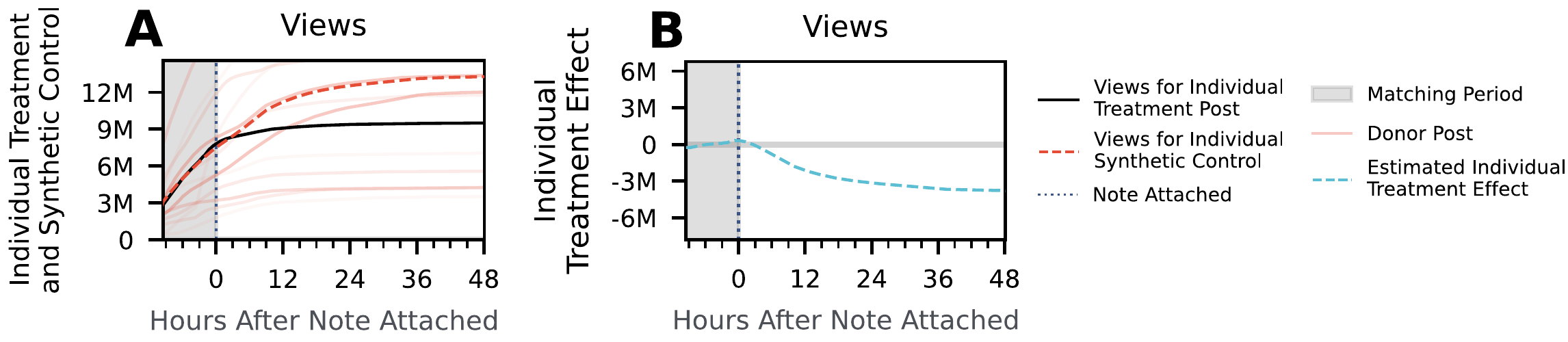}
\caption{\textbf{Illustration of synthetic control methodology for views.} \textit{(A)} Total number of views over time before and after receiving a community note for a sample post. The black line represents the post's observed view count and the dashed red line depicts the synthetic control’s view count, an estimate of what would have occurred had no note been attached. Individual donor posts that contribute to the synthetic control are shown in solid red lines with intensity of color proportional to their weight. Note that in addition to view counts the synthetic control also has a similar trajectory on all other engagement and diffusion metrics. \textit{(B)} Estimated difference between views of the synthetic control without the note and observed views with the note, quantifying the reduction in views  attributable to note attachment at a given time.}
\label{fig:1}
\end{figure*}

In this work, we investigate the causal effects of attaching community notes to posts on the engagement with and diffusion of the posts. We collect time-series data for 40,078 posts created between March and June 2023 for which community notes were proposed. We track key engagement metrics of the posts, including the number of reposts, likes, replies, and views over time. We also collect all reposts since the post was created, both before and after a note was proposed, and the follow graphs of the users that reposted them, which we use to reconstruct the diffusion cascades of the posts. These granular records concerning a post's engagement both before and after a community note appears allow us to provide precise estimates of the notes' effects, and investigate the conditions under which community notes are more or less successful at reducing the impact of misinformation.

We use synthetic control methods to estimate these causal effects~\cite{abadie_synthetic_2010, abadie_penalized_2021}. For each post with a note attached, we construct a synthetic control by averaging the engagement histories of multiple donor posts—those for whom a note was proposed, but not attached—such that the synthetic control closely matches the history of all metrics of the noted post during the period before note attachment. Then, we estimate the effect of attaching a note by comparing engagement metrics during the period after the note was attached between the post and its synthetic control. This analysis produces an individual treatment effect for each post where a community note was attached. \hyperref[fig:1]{Fig.~1}
illustrates the procedure for a sample post: \hyperref[fig:1]{Fig.~1\textit{A}} shows the number of views over time for the community noted post along with its estimated views had the note not been attached, while \hyperref[fig:1]{Fig.~1\textit{B}} shows the estimated treatment effect of the community note over time.
To validate our estimation approach, we conduct an in-time placebo test~\cite{abadie_using_2021} by artificially shifting the note attachment time one hour earlier and, as expected, estimate null effects in the period between the artificial and actual attachment time (further details can be found in~\textit{SI Appendix, section 5}).

We find that notes significantly reduce the number of views, reposts, likes, and replies. We interpret the impact on these metrics through two perspectives: a \textit{growth perspective}, which quantifies the reduction in the additional growth of the metric after a note was attached, and an \textit{overall perspective}, which quantifies the total reduction in the metric since the post's creation. The growth perspective measures effectiveness conditional on when the note was attached, while the overall perspective also accounts for the engagement that occurred before the note was attached. These varied perspectives are related to measures of the ``preventable fraction among the unexposed'' and ``preventable fraction'' in epidemiology~\cite{rothman2008modern}.

Beyond engagement, we also consider the impact of note attachment on the structure of information diffusion. Previous studies have found that fact-checked false news have larger, deeper, wider, and more viral diffusion cascades than fact-checked true news~\cite{vosoughi2018spread}. Subsequent analyses of the same data have shown that while there are significant differences in cascade size, the structural differences disappear after controlling for size, suggesting that mechanisms through which true and false fact-checked news diffuse are relatively similar~\cite{juul2021comparing}. 
In our setting, we find that note attachment qualitatively changes the structure of the post’s diffusion cascade, relative to the same post without a note. Most significantly, it reduces the depth and structural virality more than would be expected given the overall reduction in size. 

Our rich data on diverse engagement metrics (reposts, replies, likes, and views) as well as our reconstruction of diffusion cascades (enabling us to study how note attachment influences cascade structure) go beyond earlier work studying the effects of Community Notes~\cite{chuai_community-based_2024}, which considered only the effects on reposts and deletions. That prior work also used difference-in-differences methods which, unlike our synthetic control methods, rely on strong ``parallel trends'' assumptions~\cite{angrist2009mostly}. When comparable, our independent estimates also provide important corroboration of those prior estimated effects.

Since our synthetic control methods approach provides causal effect estimates at the individual post level, we can examine how average effects vary across different post subpopulations. Overall, we find that notes have the greatest absolute impact on reducing engagement when they are attached shortly after a post is created or attached to highly engaging posts. We find that notes on posts with embedded media, as opposed to text-only posts, are associated with larger reductions. We also see larger reductions on posts where concerns about altered media are presented as a reason for the note. In terms of differences across how notes are composed, we find that moderately long notes and notes written using simpler language are associated with larger reductions in engagement.

\begin{figure*}
\centering
\includegraphics[width=\linewidth]{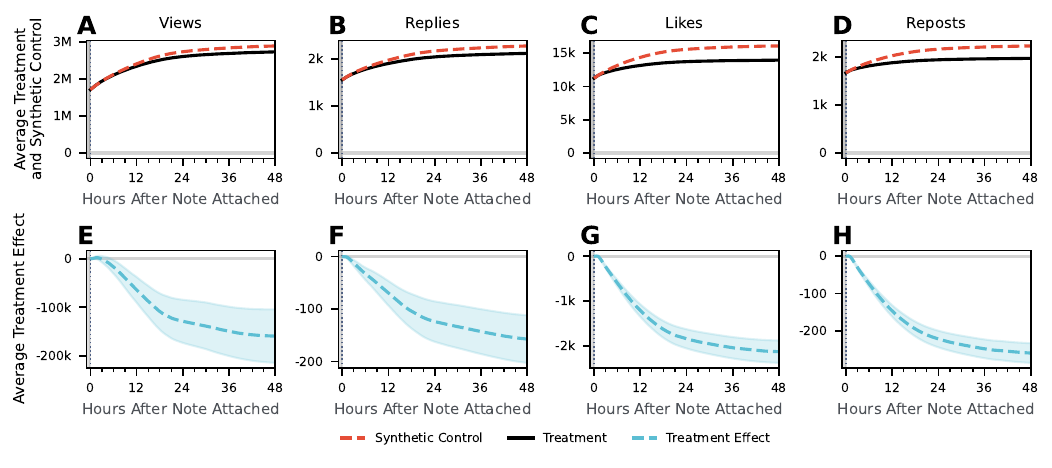}
\caption{
\textbf{Effects of note attachment on views, replies, likes, and reposts.}
\textit{(A--D)} Average treatment and synthetic control for all engagement metrics. Average treatment consists of the average value for a metric at a given time point among posts that received notes. Average synthetic control consists of the average value of the synthetic controls, estimating what the average would have been had notes not been attached.
\textit{(E--H)} Average treatment effect on the treated for all engagement metrics. Estimates the average difference between the treatment and synthetic control, i.e., the average reduction in a metric due to note attachment. The error bands represent 95\% confidence intervals.}
\label{fig:2}
\end{figure*}

\section*{Data Collection}
We collected data from March 16 to June 23, 2023, tracking 40,078 posts for which a note was proposed. We continuously monitored the ``New'' tab of the Community Notes website, which provides the identifiers of the posts for which notes were recently proposed. When a community note was created, we immediately retrieved the associated post's engagement metrics using the X API. We then made API calls every five minutes to record the post's total number of reposts, likes, replies, and views for three weeks following note creation. To ensure that all post engagement histories are comparable when constructing synthetic control weights, we shifted the engagement metrics to a timeline aligned with post creation rather than note creation time by linearly interpolating to fifteen-minute intervals. All engagement and diffusion metrics are all-cause measures, regardless of whether the content was delivered through an algorithmic or reverse-chronological feed.

We use public data available from X to determine when notes were attached to posts and which notes were never attached to any posts. Among the 40,078 posts included for analysis, 6,757 (16.9\%) received helpful notes and constituted the treatment group. (Further details on the construction of the treatment group are provided in \textit{SI Appendix, section 1B}.) The remaining 33,321 posts, for which a note was proposed but no note reached a helpful status, and thus no intervention took place, constitute the donor pool for constructing our synthetic controls. We focus our analysis on the effects of community notes within a 48-hour window after a note is attached. Given the rapid decay of engagement on X~\cite{pfeffer2023half}, our 48-hour estimates closely approximate the lifetime effects for posts that remain on the platform.

In addition to collecting engagement metrics starting when a first community note was written about a post, we also collected each post's full public repost and reply history, extending back to its creation. However, we were unable to collect this data for deleted and private posts. In such cases, we relied on the repost and reply counts returned by the X API. We collected full public repost histories for 36,408 posts (90.8\%) and full public reply histories for 30,858 posts (77.0\%). We provide further details on the data cleaning and cascade data collection in {\it SI Appendix, section 1}.

\section*{Results}
\label{sec:results}

\subsection*{Decline in Average Engagement}
\label{sec:results-engagement}

We use synthetic control methods to estimate individual treatment effects on the number of views, replies, likes, and reposts for each post with a note reaching helpful status. We then aggregate these effects into average treatment effects on the treated population. Details on the construction of synthetic controls and the uncertainty quantification underlying the confidence intervals can be found in {\it Materials and Methods}. \hyperref[fig:2]{Fig.~2\textit{A}} shows the average number of views over the 48 hours after a note was attached for posts that received community notes, along with the average number of views for the same posts' synthetic controls. \hyperref[fig:2]{Fig.~2\textit{E}} shows the average treatment effect on the treated, i.e., the average treatment effect across all noted posts. The equivalent figures for the number of replies, likes, and reposts are shown in \hyperref[fig:2]{Fig.~2~\textit{B--D}} and \hyperref[fig:2]{Fig.~2~\textit{F--H}}, respectively. These measures quantify the aggregate impact of the Community Notes program. We discuss the heterogeneities among posts in \textit{Factors Associated with Large Effects} and the distribution of individual treatment effects in \textit{SI Appendix, section 2.}

We first discuss the aggregate effect of attaching notes on the posts' number of views. We estimate that the average number of views for posts that received helpful notes decreased from 2.89 million views (95\% CI: [2.84M,~2.95M]) to 2.73 million views due to note attachment. These estimates represent the total number of views that noted posts received, or would have received without notes, 48 hours after they had community notes attached. The estimated treatment effect at this time amounts to −159,592 views (95\% CI: [−214,839, −104,344]). This corresponds to a −13.5\% reduction in additional growth in the number of views after note attachment or a −5.5\% reduction in the total number of views, including those that occurred before the note was attached. Note that the first metric captures how the presence of a note affects a post's ability to gain new views, while the second measures the overall impact of community notes at the platform level. 

While the decrease in views reflects how community notes limit the posts’ reach, change in engagement metrics that require active participation from the users---specifically replies, likes, and reposts---capture how the notes affect the way users interact with the posts. Replies on social media may signal some combination of agreement, disagreement, or confusion from a replier. We estimate that the average number of replies to posts receiving helpful notes decreased by a similar percentage as the number of views: from 2,270 replies (95\% CI: [2,225, 2,315]) to 2,112, a change of −158 replies  (95\% CI: [−203, −112]), which amounts to a −21.9\% change in reply growth after attachment or −6.9\% change in total number of replies.

Likes and reposts, on the other hand, are more frequently used as signals of positive engagement: likes on social media can indicate that the user finds the post enjoyable, useful, or interesting~\cite{meier2014more,chin_facebook_2015}, while reposts can signal agreement or serve to amplify a message~\cite{boyd2010tweet}. We estimate that the average number of likes given to posts that received helpful notes during this period fell from 16,089 (95\% CI: [15,839, 16,340]) to 13,955 due to note attachment, an absolute change of −2,134 likes (95\% CI: [−2,385, −1,884]). This amounts to a change of −44.1\% in likes after note attachment and a change of −13.3\% in total likes. Similarly, relative to the average synthetic control of 2,234 (95\% CI: [2,207, 2,260]), we estimate that note attachment led to a change of −259 reposts (95\% CI: [−285, −232]), bringing the observed average down to 1,975. This amounts to a percentage change in reposts after attachment of −46.1\% and a percentage change in total reposts of −11.6\%.

\subsection*{Altered Dynamics of Information Diffusion}
\label{sec:results-diffusion}
Having found that community notes lead to sizable reductions in average engagement, we next examine their impact on how information spreads on the platform, specifically their effect on the structure of repost cascades. A repost cascade records the tree of reposts stemming from a post. The max depth of a cascade refers to the length of the longest chain of reposts it contains, max breadth refers to the maximum number of reposts at any level of depth, and structural virality refers to the average distance between any two nodes in the cascade, standardized by its size \cite{goel_structural_2016}. A cascade with high breadth or low structural virality suggests that the post spread primarily through direct reposts of the original post. In contrast, high depth or high structural virality indicates the post spread more through multi-step, person-to-person reposting chains, a pattern often seen with rumors and viral content \cite{goel_structural_2016}. Additional details on the calculation of these metrics can be found in \textit{SI Appendix, section 1E}. 

\begin{figure*}[t]
\centering
\includegraphics[width=\linewidth]{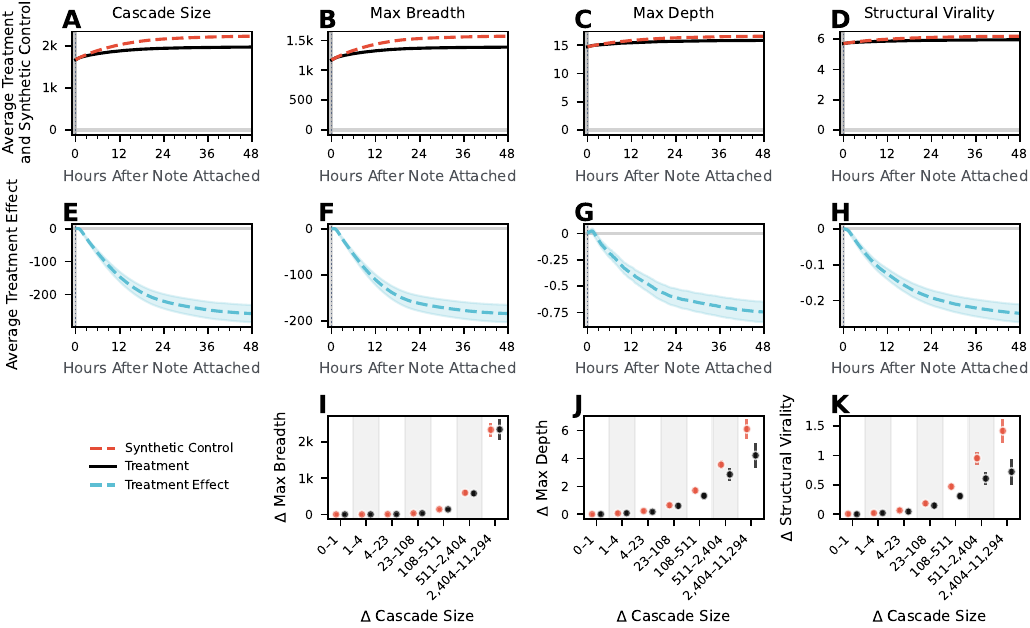}
\caption{
\textbf{Effects of note attachment on the structure of repost cascades.}
\textit{(A-D)} Average treatment and synthetic control values for \textit{(A)} the size of the repost cascade (equivalent to the total number of reposts), \textit{(B)} 
the maximum breadth of the repost cascade (which frequently occurs at the first level of the cascade and is, hence, a proxy for the number of direct reposts), \textit{(C)} the maximum depth of the repost cascade (the longest chain of person-to-person reposting), and 
\textit{(D)} the structural virality of the repost cascade (intended to measure the extent to which the post spread ``virally,'' where larger numbers indicate more viral diffusion).
\textit{(E-H)} Average treatment effect on the treated posts for all structural metrics. 
\textit{(I-K)} Change in cascade structure controlling for the growth in cascade size after note attachment. For individual treatment posts and synthetic controls whose repost cascades grew by a given amount in the 48 hours between $t=0$ and $t=48$, the plot shows the binned average growth in a structural metric over that same period. The bin edges evenly divide the positive range of $\Delta$ Cascade Size in logarithmic scale. The error bands and error bars represent 95\% confidence intervals.}
\label{fig:3}
\end{figure*}

As before, we construct a synthetic control for each post that received a helpful note, but now estimate treatment effects for structural metrics that characterize the post's repost cascade: its max breadth, max depth, and structural virality. We again take the mean of these synthetic controls, as shown in ~\hyperref[fig:3]{Fig.~3 \textit{A--D}}, along with the average observed values under treatment, and calculate average treatment effects on the treated, shown in ~\hyperref[fig:3]{Fig.~3 \textit{E--H}}. The effect on the total number of reposts, also referred to as cascade size, is repeated in \hyperref[fig:3]{Fig.~3\textit{A}} and~\hyperref[fig:3]{Fig.~3\textit{E}} for reference. 

As with engagement metrics, we estimate that note attachment leads to a reduction in all structural metrics of the repost cascade after 48 hours, relative to what would have been expected had the note not been attached. The average max breadth of synthetic controls is 1,571 reposts (95\% CI: [1,552, 1,590]), while the observed value under treatment was 1,388 reposts, producing an estimated change of −183 reposts (95\% CI: [−202, −165]) or −46.0\% in growth after note attachment, or −11.7\% in total. The estimated average max depth under control is 16.61 reposts (95\% CI: [16.51, 16.71]), compared to the observed average of 15.87 reposts, a treatment effect of −0.74 reposts (95\% CI: [−0.84, −0.65]), or −39.9\% growth after note attachment, or −4.5\% change in total. Finally, the average structural virality under control is estimated to be 6.19 (95\% CI: [6.16, 6.21]) while the observed value is 5.95, a change of −0.24 absolute units (95\% CI: [−0.26, −0.21]), or −48.5\% growth after note attachment, or −3.9\% in total.

Smaller repost cascades tend to be both less broad and less deep than larger repost cascades~\cite{juul2021comparing}. To disentangle changes in size from changes in structural metrics, we perform the following matching procedure. For each post that received a note, we first calculate the growth in observed cascade size as well as in the observed structural metrics in the 48 hours after note attachment. We apply the same procedure to the estimated synthetic controls. Finally, we compare the change in structural metrics for treated posts and synthetic controls whose repost cascades grew in size by a similar amount after note attachment. As shown in~\hyperref[fig:3]{Fig.~3 \textit{I--K}}, we find that while max breadth does not differ in the distribution between treatment and synthetic controls that grow by similar amounts, max depth and structural virality do differ. Relative to synthetic controls that grew by a similar amount, posts that receive notes do so with a smaller max depth and smaller structural virality, indicating less viral diffusion in the presence of an attached note. These results suggest that effects of attaching a community note on depth and structural virality cannot be simply explained by a change in cascade size and suggest that the attachment of a note significantly affects the mechanism through which the posts spread over the network. These structural differences are consistent with community notes having a larger moderating effect on users when the post reaches them through a repost cascade, and less of an effect on users who receive the post directly from the original poster.

\subsection*{Factors Associated with Large Effects}
\label{sec:results-hte}

\begin{figure*}
\centering
\includegraphics[width=\linewidth]{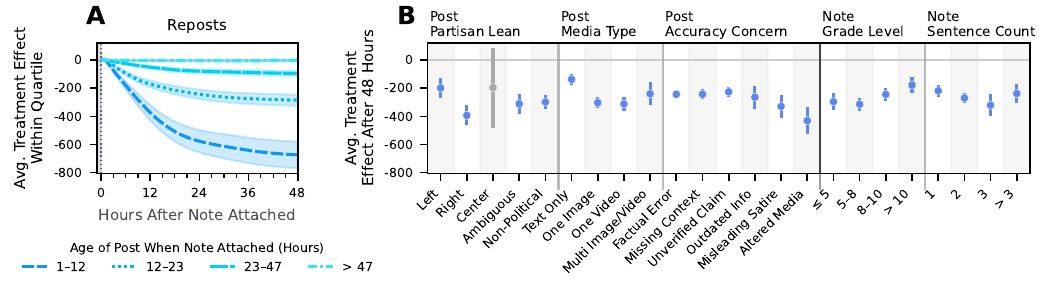}
\caption{\textbf{Factors associated with effects on reposts.} \textit{(A)} Average treatment effects for noted posts, stratified by quartiles of note attachment speed, i.e., the time elapsed between the post's creation and note attachment. \textit{(B)} Average treatment effects for noted posts after 48 hours, based on (i) the post's partisan lean, (ii) the number of images and videos, (iii) the accuracy concerns raised by the community note writer, (iv) the readability (Flesch–Kincaid grade level), and (v) the length of the community note. The error bands and error bars represent 95\% confidence intervals.}
\label{fig:4}
\end{figure*}

Our previous analyses show that community notes significantly reduce the average engagement with misleading posts and change the diffusion patterns of such posts. However, their effects are not uniform across all types of posts or notes. Next, we perform exploratory analyses to identify the contexts in which attaching community notes has the largest impact. We caution that the results presented below are associations and should not be interpreted as causal. While our synthetic controls methodology enables us to estimate the causal effects of attaching a community note to an individual post, the associations between the post or note factors and the estimated effects may be confounded by other observed and unobserved characteristics. We consider estimating the causal impact of the factors most strongly associated with large effects an important direction for future work.

In ~\hyperref[fig:4]{Fig.~4\textit{A}}, we show the average treatment effect of community notes on reposts, stratified by speed of note attachment. We find that notes attached soon after a post is created are more effective at reducing reposts. Specifically, the absolute estimated treatment effects for the first quartile (attached within 1-−12 hours), second quartile (12-−23 hours), third quartile (23-−47 hours), and fourth quartile (47+ hours) are −673 (95\%~CI: [−768,~−579]), −285 (95\%~CI: [−325,~−246]), −95 (95\%~CI: [−113,~−77]), and −2 (95\%~CI: [−13,~8]), respectively. The corresponding percentage reductions in repost growth following note attachment are −49.6\%, −44.1\%, −38.8\%, and −6.2\%, while the total repost reductions amount to −24.9\%, −12.3\%, −4.3\%, and 	
−0.1\%, respectively.

As shown in \textit{SI Appendix, section 4}, other engagement metrics exhibit similar monotonically decreasing effects when stratified by speed of note attachment. One nuance is that for posts in the fourth quartile (47+ hours from post creation to note attachment), the percentage changes in average growth of views and replies are positive: an increase of 13.6\% for views and an increase of 27.0\% for replies 48 hours after note attachment. This increase suggests that community notes may draw attention to stale posts. However, it does not imply that the notes draw more endorsement or agreement, as the average growth for likes and reposts (as opposed to views) within this quartile are −22.2\% and −6.2\%, respectively.

Another major factor associated with the magnitude of the treatment effect is the volume of engagement a post received before a note was attached: posts that received more reposts prior to note attachment have larger treatment effects on average. In \textit{SI Appendix, section 3}, we show that while the most popular posts exhibit larger absolute drops in reposts after note attachment, their percentage changes are relatively similar, suggesting these declines stem mostly from their larger baseline audience. In contrast, posts with low initial repost counts often experience small positive changes, indicating that note attachment can sometimes boost the visibility of less prominent content.

While the majority of variation in treatment effects is attributable to the speed of note attachment and the posts' popularity before treatment, we also find that both post content and note quality are associated with variability in treatment effects. Prior work has found that right-leaning media consumers tend to be more prone to motivated reasoning than left-leaning consumers~\cite{walter_fact-checking_2020} and that Republicans are less likely than Democrats to rate community notes as helpful~\cite{wojcik_birdwatch_2022}. These findings suggest that notes on right-leaning content, likely seen more often by right-leaning consumers~\cite{cinelli_echo_2021}, would have smaller treatment effects than those on left-leaning content. However, we do not find evidence that this is the case among the English-language posts that we label for partisanship. In fact, we estimate that notes on left-leaning content tend to be less effective at reducing reposts, both in absolute and relative terms. We describe the methodology for labeling post partisanship in \textit{SI Appendix, section 1F}. The estimated average treatment effect on right-leaning posts is −393 reposts (95\% CI: [−464, −323]), a percentage change in average growth due to note attachment of −55.1\%, relative to an estimated average treatment effect on left-leaning posts of −198 reposts (95\% CI: [−269, −128]), a percentage change of −41.8\%. As shown in \textit{SI Appendix, section 4}, we also find similar effects for views, replies, and likes. 

Beyond partisanship, we also find that the type of media included in the post is associated with note effectiveness. The estimated treatment effect for media posts—those containing a single image, single video, or multiple images/videos—are −303 reposts (95\% CI: [−343, −263]), −312 reposts (95\% CI: [−365, −260]), and −240 reposts (95\% CI: [−322, −158]), respectively. In contrast, text-only posts have an estimated treatment effect of −137 (95\% CI: [−175, −98]). While media posts generally receive more engagement than text-only posts before receiving a note, we find that the difference in average effect persists even after normalizing by size. The estimated percent changes in average growth due to note attachment are −50.7\%, −45.6\%, −48.7\% for a single image, single video, and multi-image/video posts, respectively, compared to −37.9\% for text-only posts. 

The sizes of the treatment effects also vary depending on what type of concern the community note addresses. When proposing a community note, writers are asked to indicate their accuracy concern, i.e., what aspect of the post they consider misleading. They can select multiple concerns, e.g., that the post both contains outdated information and makes a factual error. As shown in ~\hyperref[fig:4]{Fig.~4\textit{B}}, we find that concerns related to altered media and misleading satire are most strongly associated with large effects. 

Considering the community notes themselves, we find that both readability and length are associated with effectiveness. We measure readability using the Flesch-Kincaid grade level score, an estimate of the minimum U.S.\ grade level required to comprehend a text~\cite{kincaid1975derivation}. We find that simpler notes tend to be more effective. Notes with grade levels less than or equal to five have an estimated average treatment effect of 	−296 reposts (95\% CI: [−356, −235]), corresponding to a −47.4\% change in average growth after note attachment. In contrast, notes with a grade level above ten have an estimated average treatment effect of −178 reposts (95\% CI: [−234, −121]), which amounts to a change of −38.8\%. Finally, we find that moderately long notes (two or three sentences) are slightly more effective than both short (one sentence) and long (more than three sentences) notes at reducing reposts.

\section*{Discussion}
\label{sec:discussion}
As the problem of misinformation persists on social media, scalable interventions are necessary to prevent its spread, impact, and harm. Crowd-sourced fact-checking presents one such approach, which has demonstrated encouraging results in early tests and is now deployed as a core content moderation component on one of the world's largest social media platforms. The public nature of Community Notes' deployment on X, with freely available source code and rating data, as well as detailed information on the content and engagement of posts over time, has allowed us to estimate the impact of community-driven fact-checking through a completely independent audit. 

Our results indicate that once community notes are attached, on average, they reduce the engagement with and diffusion of false information on X. Consistent with related work by Chuai et al.~\cite{chuai_community-based_2024}, which studied reposts and deletions, we find that community notes lead to a decline in the number of reposts that a post receives after attachment. Despite differences in causal identification strategies, our estimate of percentage decline in average reposts after note attachment, −46.1\% during the period March 16 to June 23, 2023, is roughly comparable with estimates from Chuai et al.\ of −55.2\%, −49.6\%, −45.6\%, and −47.5\% in the months of March, April, May, and June of 2023. Our analysis of other outcomes finds that notes lead to a similar percentage reduction in likes (−44.1\%) but smaller reductions in views (−13.5\%) and replies (−21.9\%). 

These findings suggest that the impact of note attachment is strongest on public expressions of support for content (reposts and likes), while its effect is smaller on whether content reaches people in the first place (views) or whether they choose to engage in an online conversation about it (replies). While these differences might be taken to imply that the reduced support nullifies the effects of misinformation exposure after attachment, we caution against this interpretation. Viewing false information, even if the viewer initially doubts its validity, can increase their likelihood of agreeing with it later \cite{pennycook_prior_2018}. Thus, each view prevented by a community note is meaningful. We also note that the decline in reposts and likes may not necessarily reflect a decline in actual support of the content but rather a reduced willingness among users to signal their support publicly. An internal analysis by X does report that users are less likely to agree with the substance of potentially misleading posts when presented with a community note~\cite{wojcik_birdwatch_2022}.

The decline in engagement is paired with notable structural changes in how posts diffuse across the platform. 
The changes are consistent with larger behavioral changes by users who do not follow the original post’s author. Such larger changes can be attributed to homophily \cite{mcpherson_birds_2001}, where users closer to the root author may share similar beliefs or information evaluation approaches with the root author. As an alternative mechanism, these results are also consistent with dyadic social pressures~\cite{boyd2010tweet} whereby users may feel greater loyalty obligations to close connections than those encountered through deeper network~paths.

While we find that community notes effectively reduce engagement once attached (−13.5\% views, −21.9\% replies,  −44.1\% likes, and −46.1\% reposts), we also find evidence that the system would be much more effective if notes were attached faster. Moreover, the reductions in views, replies, likes, and reposts are much more modest when measured as a percentage of overall engagement with the post, compared to only considering changes in engagement after note attachment. When measured this way, the overall percentage changes in views, replies, likes, and reposts due to note attachment are −5.5\%, −6.9\%, −13.3\%, and −11.6\%, respectively. 

Our study has several limitations worth considering. 
First, like most observational studies, our analysis relies on certain assumptions to identify causal effects. To examine the plausibility of these assumptions, we perform a series of robustness checks, including (i) a placebo test, (ii) limiting the donor pool for constructing synthetic controls to posts with notes that received high scores from the Community Notes algorithm, and (iii) incorporating semantic embeddings of the posts---alongside engagement and diffusion trajectories---when constructing synthetic controls. The results of the placebo test were consistent with suitable controls being in place (\textit{SI Appendix, section 5}), and neither limiting the donor pool nor incorporating the posts’ semantic embeddings into the construction of the synthetic controls substantively changed the conclusions of our analysis (\textit{SI Appendix, sections 7 \& 8}).
Second, our analysis can only estimate the effects of community notes on posts that had notes attached. Based on the data available to us, we cannot estimate the coverage of the Community Notes program (i.e., how many misleading posts on X do or do not received a note) or what the effectiveness of notes might be on the broader population of posts without notes. Such analysis is inherently challenging as it requires both access to all posts on the platform and a scalable method for identifying misleading posts.
Third, we cannot test for implied truth effects~\cite{pennycook_implied_2020}, i.e., the potentially increased tendency for posts without community notes to be perceived as accurate or non-misleading, even when they are~not.
Finally, Community Notes is an evolving system and our analysis reflects the effects of the system during the study period, March--June 2023. Like all social media research, the rapidly changing environment makes temporal validity challenging~\cite{munger2019limited}. Since we concluded our data collection, more volunteers have joined the program, extensions of the system have been proposed~\cite{de2025supernotes}, and substantial updates to the system’s implementation have been introduced. These updates include improvements in the time required to run the algorithm and display the notes \cite{communitynotes_ranking_2024}, as well as the automatic attachment of a community note to posts with images and links that were previously included in other noted posts. Our findings suggest that these changes are likely to lead to significant additional reductions in engagement with misleading content on X. Nevertheless, community-based fact-checking is best viewed as one of several interventions~\cite{bak2022combining} worth considering when aiming to reduce the spread of misinformation on social~media.

\section*{Materials and Methods}
\label{sec:methods}

\subsection*{Estimating Individual Treatment Effects}
We use synthetic control methods to estimate the effect of receiving a community note on a post's engagement and diffusion. These methods are commonly employed to estimate causal effects when detailed time-series data is available for both treated and untreated units~\cite{abadie_synthetic_2010, abadie_penalized_2021}. Under the synthetic controls framework, each unit that receives the intervention (in our case, note attachment, which we also refer to as ``treatment'') receives an individual synthetic control estimate, which is interpreted as what would have happened to the treated unit had the intervention not occurred (i.e., had it never received a community note). Observations of the outcome for treatment unit $i$ are denoted as $Y_{imt}(Z_i=1)$, where $m$ refers to the metric in question (e.g., reposts), $t$ refers to the amount of time elapsed since note attachment (e.g., 48 hours after attachment), and the value of $Z_i=1$ indicates that we are referring to unit $i$ when it received treatment. We refer to estimates from the synthetic control as $\hat Y_{imt}(Z_i=0)$.

A synthetic control is a weighted average of donor posts: $\YhatCimt = \sum_{j \in D_i} w_{ij} \YCjmt$, where $D_i$ is the set of eligible donors for post $i$ and $w_{ij}$ are non-negative weights. As donors, we considered posts for which at least one community note was proposed, but no note was ever found to be helpful, and which therefore never experienced note attachment. For each treated post, we further restricted the donor pool to untreated posts for which historical data was available in the time period around which the treated post received a helpful note. For treated posts with more than 12 hours of historical data, we used only the 12 hours immediately preceding treatment. This choice allowed us to avoid requiring donor posts to always have histories as long as, or longer than, those of the treated posts, thus helping us maintain a suitable number of donors for each treated post. To ensure a sufficient amount of data for constructing estimates for each individual metric, we considered only treated posts with at least one hour of historical data prior to note attachment for the given metric.

To determine the weights for each synthetic control, we minimized the squared Euclidean distance between the metrics of the treated post and those of its synthetic control, using data up to the time point when the treated post had a note attached:
$\sum_{(m,t) \in M_i \times T_i : t < 0} {[Y_{imt}(1) - \hat Y_{imt}(0)}]^2$, 
where $M_i$ refers to the complete set of metrics for unit $i$, and $T_i$ refers to the complete set of time points. We construct weights by minimizing this distance across the following metrics, when available for a post: views, replies, likes, reposts, author follower count, repost cascade maximum breadth, repost cascade maximum depth, and repost cascade structural virality (\textit{SI Appendix, section 1D}). In total, inferring the weights required solving 6,757 linearly constrained least squares problems (one for each treated post), each of which was a quadratic program. Due to the computational burden of solving each of these programs, we restrict the donor pool for each treated post to the 1,000 control posts closest to that treated unit in Euclidean distance. This restriction greatly reduces the computational cost of solving each quadratic program, and can be viewed as a hard thresholding analog of the ``penalized'' synthetic control method \cite{abadie_penalized_2021}. 
To prevent metrics with larger scales (e.g., views) from dominating the synthetic control construction, we standardize all metrics by their sample standard deviation within the treated posts, following Abadie \& L’Hour \cite{abadie_penalized_2021}.

We refer to the true individual treatment effect for unit $i$, metric $m$, at time point $t$ as $\tau_{imt}=Y_{imt}(1) - Y_{imt}(0)$, and $\hat\tau_{imt}$ as our estimate of that quantity using synthetic controls. One approach to estimating this effect would be to consider the simple difference between the treated post's engagement and the corresponding engagement for its synthetic control, $Y_{imt}(1) - \hat Y_{imt}(0)$. However, this estimation method can induce bias in $\hat\tau_{imt}$ if the treatment unit and synthetic control do not closely match prior to treatment, which is not always possible in high-dimensional datasets~\cite{ben-michael_augmented_2021}. We therefore employ the bias correction procedure recommended by Abadie \& L’Hour~\cite{abadie_penalized_2021} to address imperfect matches between treatment posts and their synthetic controls when estimating individual treatment effects. 
The procedure involves fitting regression models on untreated posts that predict all post-treatment outcomes (one model for each metric and time point combination) based on their pre-treatment history. We fit ordinary least squares models. To build the training datasets, we sampled 100 donor posts from each treated unit’s donor pool and combined them. We then estimate the bias-corrected synthetic controls as:
$\YhatCimtBC = 
    \sum_{j \in D_i} w_{ij} \YCjmt - 
    \sum_{j \in D_i} w_{ij} [\Mmt (\boldsymbol{X_i}) - \Mmt (\boldsymbol{X_j})]$,
where $\Mmt$ is the model for metric $m$ for post-treatment time period $t$, and $\boldsymbol{X_i}$ and $\boldsymbol{X_j}$ are the pre-treatment covariates.

\subsection*{Aggregating Individual Treatment Effects}
We calculate several statistics to summarize the estimated individual effects. First, we calculate the average estimated treatment effect on the treated posts as $\hat{\tau}_{\cdot mt}=\frac{1}{|N_{\cdot mt}|}\sum_{i \in N_{\cdot mt}}{\hat{\tau}_{imt}}$ (shown in \hyperref[fig:2]{Fig. 2 \textit{E--H}} and \hyperref[fig:3]{Fig. 3 \textit{E--H}}), where $N_{\cdot mt}$ refers to the complete set of treated units for a metric $m$ and time point $t$. Referring to the average observed value under treatment as $\YTdotmt$, and the average estimated (bias-corrected) synthetic control value as $\YhatCdotmtBC$, then $\hat{\tau}_{\cdot mt}$ is equivalent to $\YTdotmt - \YhatCdotmtBC$. This means that the average estimated treatment effect can be interpreted as the absolute change in average outcome due to note attachment.

To quantify the uncertainty of the synthetic control estimation procedure, we use standard 95\% Gaussian confidence intervals, defined as: $\hat{\tau}_{\cdot mt} \pm z_{\alpha/2} \frac{\hat{\sigma}_{\cdot mt}}{{|N_{\cdot mt}|}}$, where $\hat{\sigma}_{\cdot mt}$ denotes the estimated standard deviation of the treatment effect for a given metric and time point. In the analyses of factors associated with large effects, we use the equivalent confidence intervals, restricted to the relevant subpopulations. In addition to reporting confidence intervals, we conduct a permutation test to further assess the statistical significance of our results (\textit{SI Appendix, section 6}).

In addition to calculating absolute change, we also compute the percentage change in average outcome due to note attachment: $\frac{\YTdotmt - \YhatCdotmtBC}{\YhatCdotmtBC}$. This metric normalizes by the average estimated outcome had notes never been attached to the posts considered, reflecting the overall impact of the Community Notes program. Importantly, it also accounts for the time required for notes to be written and rated as part of the treatment. To summarize the effect of notes once they have been attached, we also consider the growth after treatment: $\Delta \YTdotmt = \YTdotmt - Y_{\cdot m0}(1)$ in the treatment unit, along with the comparable growth in the control unit, $\Delta \YhatCdotmtBC = \YhatCdotmtBC - \hat Y^{BC}_{\cdot m0}(0)$. The percentage change in growth is then $\frac{\Delta \YTdotmt - \Delta \hat{Y}^{BC}_{\cdot mt}(0)}{\Delta \hat{Y}^{BC}_{\cdot mt}(0)}$.

\section*{Acknowledgments}
We thank Jennifer Allen, Michael Bernstein, Kayla Duskin, Dean Eckles, Andy Haupt, Jonas Sch{\"o}ne, and Amy X. Zhang for helpful feedback and discussions. This work was supported in part by a University of Washington Information School Strategic Research Fund Award, cloud computing credits by Google, and Army Research Office Multidisciplinary University Research Initiative (MURI) Award W911NF−20−1−0252.

\begingroup
\small
\setlength{\bibsep}{4pt}
\bibliographystyle{unsrt}
\bibliography{citations}
\endgroup

\clearpage

\appendix
\renewcommand{\thesection}{\arabic{section}}

\renewcommand{\thesubsection}{\Alph{subsection}}

\setcounter{figure}{0}
\renewcommand{\thefigure}{S\arabic{figure}}

\setcounter{table}{0}
\renewcommand{\thetable}{S\arabic{table}}

\section*{{\LARGE Appendix}}

\section{Additional Details Concerning Data Collection and Processing}
\label{si:data-collection}

\subsection{Data Collection Pipeline}
\label{si:data-collection-pipeline}
We collected data from March 16 to June 23, 2023. On June 23, 2023, the X Academic Research API was discontinued, and a new pricing structure was implemented, making further data collection prohibitively expensive. We initially collected data for 41,310 posts that were the subject of community notes, using the \url{/2/tweets} X API endpoint to collect counts of views, replies, likes, and reposts over time. We started querying this endpoint soon after a note for a given post was proposed and appeared on the ``New'' tab of the Community Notes website, which provides the identifiers of posts for which notes were recently proposed. Since the ``New'' tab includes only posts that have been liked or reposted more than 100 times in total, we also scanned the most recently released publicly available Community Notes data every hour to ensure we did not miss any posts. We continued querying the \url{/2/tweets} endpoint every five minutes for the following three weeks. 

During data processing, we identified anomalies in the engagement metrics returned by the API for 874 of these posts (2.1\%). These anomalies consisted of sharp rises and declines in engagement metrics (usually in only a single engagement metric), which we hypothesize were due to data processing issues on the X backend. We removed posts exhibiting such anomalies from our dataset, resulting in the inclusion of 40,436 posts after filtering. We provide more details concerning the identification of anomalous posts in Section~\hyperref[sec:apifiltering]{\textit{1.C}}.

We applied two additional filters: posts needed at least one hour of data prior to note attachment to ensure reliable synthetic control construction, and at least 48 hours of data afterward to maintain consistency in the set of posts used to compute average treatment effects across time points. These criteria excluded 355 posts (0.9\%), reducing the dataset size to 40,081. Three treated posts were later excluded because the quadratic programs used to determine their synthetic control weights failed to converge, resulting in a final count of 40,078.

In addition to collecting data through the \url{/2/tweets} endpoint, we also used the full-archive (\url{/2/tweets/search/all}), recent search (\url{/2/tweets/search/recent}), and the follow graph (\url{/2/users/:id/followers} and \url{/2/users/:id/following}) endpoints. These endpoints provide information on which accounts publicly reposted or replied to posts that received community notes, allowing us both to calculate the exact number of reposts and replies at any given time since the posts’ creation and to construct the repost cascades. Since these counts were exact, we used them as our default repost and reply count metrics, falling back on the repost and reply counts returned by the \url{/2/tweets} endpoint when they were unavailable.

\subsection{Defining Treatment Status}
X runs the ``bridging-based'' matrix factorization algorithm that classifies community notes as ``helpful'' every hour. While the majority of notes classified as ``helpful'' remain ``helpful'' for the 48-hour period over which we analyze the effects of the notes, some notes lose their ``helpful'' status. In our analysis, if a post had a ``helpful'' note at any point, even if it was later reclassified as ``needs more ratings'' or ``not helpful,'' we consider it part of the treatment group for its entire lifespan. Of the 6,757 treated posts in our analysis, 5,366 had a ``helpful'' note for the entire 48-hour period after first receiving one (79.4\%), meaning that 20.6\% of posts experienced at least one additional change in status.

Transitions from ``helpful'' to ``needs more ratings'' or ``not helpful''  tend to occur after many viewers have already seen a post. The median transition occurred 15.5 hours after posts were originally rated ``helpful.'' In fact, 65.9\% of views for posts with notes that lost ``helpful'' status occurred while the post had a ``helpful'' note. Together, these facts indicate that most views of treated posts occur when the posts have notes attached: among all treated posts, 93.4\% of views in the 48 hours after note attachment occurred while the posts had ``helpful'' notes.

Only one note can be attached to a post at a time, and when multiple notes are classified as ``helpful'' for a given post, X randomly selects one to display. Of the 6,757 treated posts, 852 had more than one note classified as ``helpful'' in the 48 hours after initial note attachment (12.6\%). For posts with multiple ``helpful'' notes, we use the note that first received a ``helpful'' status to determine the treatment time. Considering all treated posts, there were a total of 7,756 notes potentially shown with the 6,757 treated posts in the 48 hours following the first note attachment. 

Data was not available on which note was actually shown alongside a post at a given time. When calculating post-level statistics related to the note a post received---e.g., a note's reading grade level in \textit{Main Text, Fig.~4}---we take a weighted average across the different notes that appeared with the post, with weights proportional to the amount of time each individual note was rated as ``helpful'' during the 48 hours after note attachment.

\subsection{Anomalous Post Removal}
\label{sec:apifiltering}
While exploring our dataset, we identified several posts that we considered to be anomalous. These posts showed large rises and drops in a subset of their engagement metrics, while other metrics exhibited more gradual changes over the same time period. In most cases only a single engagement metric was affected. 
We did not believe the metrics returned by the X API accurately reflected the true engagement with these posts and therefore elected to remove them from the analysis. 

To identify posts for removal, we began by plotting a random sample of 300 posts that showed more than a 1\% rise or drop in at least one metric between API calls, which must have also amounted to at least 10 absolute units, a permissive heuristic that captured all posts identified as anomalous during initial checks. Two authors (I.S. and M.S.) each labeled the engagement metrics of 150 posts as either (i) accurate or (ii) likely erroneous and in need of removal. After discussing the labeling criteria, they conducted a second round of labeling on an additional random sample of 300 posts that exhibited a drop of 0.5\% to 3\% in a metric (amounting to 20 to 100 absolute units), combined with a rise of the same magnitude. The authors each labeled 200 of the 300 posts and gave the same label to 49 of the 50 overlapping records (Krippendorff's $\alpha=0.96$, 95\% CI: [$0.85, 1$]).

After labeling a total of 600 posts, we tested various criteria for identifying anomalous posts. We performed a grid search to identify combinations of percentage and absolute rises and drops that minimized mislabeling. From the grid search, we selected the thresholds that maximized the number of posts that were correctly removed, among solutions that did not incorrectly remove \textit{any} posts from the labeled training data. The optimized thresholds were a rise of 25 units, amounting to at least 3\%, and at another point in time, a drop of 25 units, amounting to at least 3\%. We finally applied these thresholds to the complete dataset, removing a total of 874 posts.

\subsection{Missing Data}
\label{sec:data-missingness}
As described above, in addition to the publicly released Community Notes data, we also collected data through multiple X API endpoints. While we were able to collect complete data for most posts, there were cases where we could retrieve data from one endpoint but not another, or where we could retrieve data from an endpoint but only for some metrics (e.g.,  replies, likes, and reposts present, but views missing). Since we requested data every five minutes and later linearly interpolated observations to 15-minute intervals from each post's creation time, this missing data mostly did not pose an issue for the analysis. However, for some posts, we were never able to retrieve data from certain endpoints, or we found that the data we did retrieve consistently lacked observations for a specific metric. We also found cases where a single endpoint would stop returning responses after a certain time, while other endpoints would continue returning responses. Similarly, we observed that some metrics would stop being returned at a given time, while others would not. Rather than removing these posts from the analysis, we decided to use only the metrics that were available. For example, for posts missing view observations, we did not include views in the construction of synthetic controls and did not estimate the treatment effects on views. In Tables \hyperref[tab:s1]{\textit{S1}} and \hyperref[tab:s2]{\textit{S2}}, we report the number of posts for which each metric was available.

\begin{table}[t]
\begin{center}
\scriptsize
\begin{tabular}{l | r r | r r}
\toprule
  & \multicolumn{2}{c|}{Control} & \multicolumn{2}{c}{Treatment} \\
  & Unavailable & Available & Unavailable & Available \\
\midrule
Reposts & 110 & 33,211 & 16 & 6,741 \\
Replies & 122 & 33,199 & 10 & 6,747 \\
Likes & 740 & 32,581 & 208 & 6,549 \\
Views & 1,664 & 31,657 & 373 & 6,384 \\
Repost Cascade Depth & 10,088 & 23,233 & 1,171 & 5,586 \\
Repost Cascade Width & 10,088 & 23,233 & 1,171 & 5,586 \\
Repost Cascade Structural Virality & 10,365 & 22,956 & 1,176 & 5,581 \\
\bottomrule
\end{tabular}
\caption{\textbf{Missing data by metric and treatment status}. Number of posts with at least one observation available for a metric, compared to the number of posts for which no observations could be retrieved.}
\label{tab:s1}
\end{center}
\end{table}

\begin{table}[t]
\begin{center}
\scriptsize
\begin{tabular}{l | r r r r r}
\toprule
& \multicolumn{5}{c}{Treatment} \\
& Fully Available & Dropped Pre Treatment & Dropped Post Treatment & Only Post Treatment & Unavailable \\
\midrule
Reposts & 6,416 & 49 & 261 & 15 & 16 \\
Replies & 6,631 & 1 & 6 & 109 & 10 \\
Likes & 4,959 & 70 & 375 & 1,145 & 208 \\
Views & 4,842 & 69 & 366 & 1,107 & 373 \\
Repost Cascade Depth & 4,264 & 407 & 915 & 0 & 1,171 \\
Repost Cascade Width & 4,264 & 407 & 915 & 0 & 1,171 \\
Repost Cascade Wiener Index & 4,259 & 405 & 898 & 19 & 1,176 \\
\bottomrule
\end{tabular}
\caption{\textbf{Amount of data available for treatment posts}. For each metric, the number of posts that had observations (i) available for at least 48 hours after treatment, (ii) available at some point before treatment but dropped before treatment occurred, (iii) available at time points both before and after treatment but not for a full 48 hours, (iv) available only after treatment had already occurred, and (v) never available.}
\label{tab:s2}
\end{center}
\end{table}

\subsection{Structural Characteristics of Repost Cascades}
For a given post, we construct its repost cascade as a directed tree using time-inferred diffusion~\cite{goel_structural_2016}, a standard procedure for attributing diffusion pathways~\cite{vosoughi2018spread}. The post itself is the root node, and each repost constitutes another node in the cascade. Each repost has a directed edge to the likely source through which the reposter was exposed to the post, according to the following procedure: For each repost, we scan the set of users the reposter follows to find the one who most recently shared the content. A directed edge is then added from the reposter’s node to the followee’s node. If no user the reposter follows has reposted the content previously, a directed edge is added to the root node.

The maximum depth of a repost cascade refers to the longest path between the original post and any individual repost in the cascade graph. The maximum breadth is defined as the largest number of reposts occurring at any single level of depth. In addition to depth and breadth, we also calculate the structural virality of a post’s repost cascade, which is closely related to the Wiener index~\cite{wiener1947structural} and captures the extent to which a post diffused virally (person-to-person) versus through a large broadcast. Structural virality is calculated as the average distance between all pairs of nodes in the repost cascade, treated as undirected~\cite{goel_structural_2016}. We calculate the exact values of maximum depth, maximum breadth, and structural virality every 15 minutes after a post was created, based on the timestamps of reposts provided by the API. Since repost data was generally collected after the fact, deleted reposts are not included in the cascades. Reconstructing repost cascades requires data that we were not able to collect for all posts; as such, we only analyzed cascades for posts for which this data was available (Section~\hyperref[sec:data-missingness]{\textit{1.D}}).

\subsection{Post Partisanship Classification}
We used the Claude 3.5-Sonnet V2 model (``claude−3−5-sonnet-v2@20241022'') to label the partisanship of English-language posts. The prompt used was ``Please annotate the tweet below using the following schema. It is okay if you cannot view photos, videos, or links from the tweet, but please do your best to interpret the text and any available context. Ensure that your response ends with a correctly formatted JSON containing all requested columns. Make sure that you do not include any other text after the JSON.'' A full JSON schema was provided, in which the description of the partisanship class was given as ``The political leaning of the tweet content. The category `unknown' indicates the political leaning is unclear, and `none' indicates the content is not political.'' (In plots, ``unknown'' was changed to ``ambiguous'' and ``none'' was changed to ``non-political'' for clarity.) There were 904 posts labeled as ``left,''
1,360 posts labeled as ``right,''
53 posts labeled as ``center,''
691 posts labeled as ``unknown,''
1,908 labeled as ``none,''
1,580 non-English posts not labeled, and
261 posts not labeled due to unavailable post text. 

\section{Distribution of Treatment Effects}
\begin{figure*}[t]
\centering
\includegraphics[width=\linewidth]{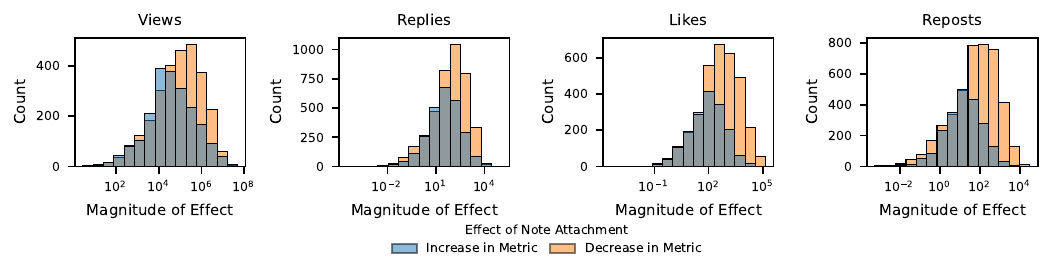}
\caption{\textbf{Distribution of individual treatment effects on each engagement metric}. Color indicates whether the treatment effect was positive (i.e., the note led to an increase in the metric) or negative, while the position on the x-axis (log scale) represents the magnitude of the increase or decrease. Plots show treatment effects after 48 hours.}
\label{fig:s4}
\end{figure*}

While we estimate that community notes result in declines in engagement on average, we find that for many posts, note attachment leads to \textit{increased} engagement relative to their synthetic controls. We estimate that 	43.0\% of posts see increases in views after 48 hours due to note attachment, while the comparable number for replies is 39.2\%, for likes is 33.9\%, and for reposts is 33.0\%. However, the estimated increases are typically smaller in magnitude than the decreases, resulting in the sizable negative average effects we observe. We plot the distributions of individual treatment effects for these four metrics in \hyperref[fig:s4]{\textit{Fig.~S1}}, showing increases and decreases separately. The plots show histograms of treatment effect magnitudes on a common log-scale axis, where the color indicates whether the treatment effect was positive or negative.

Considering positive and negative treatment effects separately, the average effects are 518,882 and −671,893 for views, respectively; 328 and −470 for replies; 799 and −3,641 for likes; and 120 and −445 for reposts. The medians show similar differences: the median positive treatment effect for views is 34,524, for example, about three times smaller in magnitude than the median negative treatment effect of −109,949. For replies, the medians are 39.3 and −97.8; for likes, the medians are 105 and −467; and for reposts, the medians are 16.4 and −75.5. (Medians are non-integer as synthetic control estimates need not be integers.) This analysis suggests that while community notes do at times result in increases in engagement, the increases tend to be both less frequent and smaller in magnitude than the decreases.

We state in the main text (\textit{Decline in Average Engagement}) that the average declines in views and replies after note attachment are −13.5\% and −21.9\%, respectively, compared to −44.1\% and −46.1\% for likes and reposts. The difference between these sets of metrics suggests that community notes may have a stronger impact on engagement that clearly signals support for a post than on the number of people who simply view it or who engage with it in a more ambiguous way. The percentage of treatment effects that are positive (43.0\% and 39.2\% for views and replies, compared to 33.9\% and 33.0\% for likes and reposts) suggests that part of this effect is because community notes more often lead to increases in views and replies than in likes and reposts. The distributions of magnitudes for positive and negative effects suggest another dimension to the differences in averages. When comparing positive and negative treatment effects that come from the same percentile of the magnitude distribution, we find that for views and replies, the positive and negative treatment effects tend to be more similar to each other than the corresponding percentiles for likes and reposts. For example, the median negative views and replies treatment effects are 3.18 and 2.49 times as large as their positive counterparts, respectively. In comparison, the median negative likes and reposts treatment effects are 4.46 and 4.61 times as large as their positive counterparts. When considering the complete set of percentiles between the median and the 99th, we in fact find that the maximum ratios of negative to positive treatment effect (i.e., how much larger the negative effect is than the equivalent positive effect) are 3.26 and 2.63 for views and replies, compared to maximum ratios of 6.27 and 6.14 for likes and reposts. The \textit{minimum} ratios among these percentiles are in fact 4.46 and 3.01 for likes and reposts, compared to 0.934 and 0.818 for views and replies. These findings suggests that, beyond positive signals of engagement being more likely than non-positive signals to decrease due to note attachment, when they do decrease, they also tend to do so by a relatively larger amount.

\begin{figure*}[t]
\centering
\includegraphics[width=0.98\linewidth]{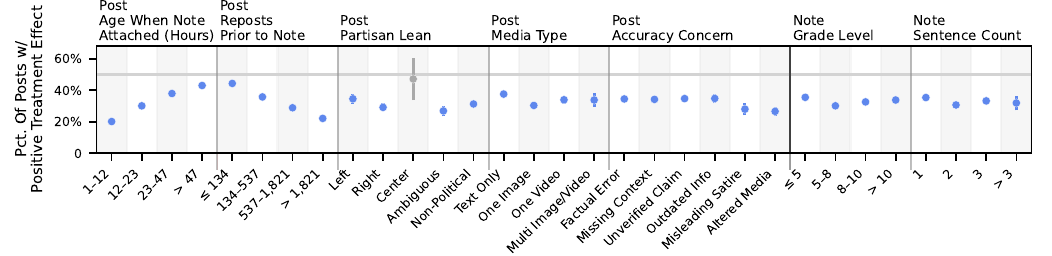}
\caption{\textbf{Factors associated with positive treatment effects on reposts}. Among all treated posts, the percentage with positive individual treatment effects after 48 hours, broken down by (i) hours from post creation to note attachment, (ii) number of reposts at the time of note attachment, (iii) post partisanship, (iv) number of images and videos, (v) accuracy concern raised by the community note writer, (vi) readability (Flesch--Kincaid grade level), and (vii) length of the community note. 
} 
\label{fig:s5}
\end{figure*}

Considering reposts, as we do in \textit{Main Text, Fig.~4}, we find that many factors associated with differences in average treatment effects (e.g., the amount of time between post creation and note attachment) are also associated with posts' probabilities of having a positive treatment effect. In \hyperref[fig:s5]{\textit{Fig.~S2}}, we plot the percentage of treatment effects that are positive, disaggregated by variables previously found to be associated with differences in treatment effects. The age of a post when it receives a note has one of the stronger pairwise associations with whether attachment will lead to an increase in reposts: 20.2\% of posts in the most rapid quartile (those noted within 12 hours of creation) have positive treatment effects, compared to 43.0\% of posts in the final quartile (those noted after 47 hours). We discuss the heterogeneity of effects by popularity, i.e., number of reposts prior to note attachment in the next section. Partisan lean, media type, accuracy concern, grade level, and sentence count show similar, if less strong, relationships with the positivity of the reposts treatment effects as they do with its magnitude.

\section{Heterogeneity in Effects on Reposts Based on Popularity Before Note Attachment}
As noted in Section \textit{Factors Associated with Large Effects} in the main text, we find that a post's popularity prior to receiving a community note is associated with the absolute treatment effect it receives. As in the main text, we center our analysis on reposts. We find that for posts with the fewest reposts prior to note attachment (i.e., those in the first quartile, having 134 or fewer reposts), the mean absolute treatment effect after 48 hours was a change of 1 repost (95\% CI: [−10, 13]), or 1.4\% overall. In comparison, the absolute changes for the second (134-−537 reposts), third (537-−1,821 reposts), and fourth (1,821+ reposts) quartiles were −67 (95\% CI: [−79, −54]), −231 (95\% CI: [−264, −198]), and −753 (95\% CI: [−849, −657]), respectively, corresponding to percentage changes of −15.0\%, −15.0\%, and −10.9\% overall.

The similarity in percentage changes between posts in the second, third, and fourth quartiles suggests that much of the difference in absolute changes (−67, −231, and −753) may be due simply to the number of users who see a post and can therefore be affected by note attachment. In other words, community notes on posts of a sufficiently large size have relatively similar effects on individual viewers' reposting behavior. However, community notes attached to posts in the smallest popularity quartile do appear to have a different aggregate effect on reposting behavior. As shown in \hyperref[fig:s5]{\textit{Fig.~S2}}, posts in this quartile are more likely to have notes lead to increases in reposts: 44.4\% of posts in the bottom quartile have positive treatment effects, compared to 35.8\%, 28.9\%, and 22.2\% of posts in the larger quartiles. These findings may indicate that attaching notes sometimes draws additional attention to low-visibility posts, resulting in a higher likelihood of positive changes. With that said, we also note that the coefficient of variation is much larger in magnitude for small posts: 190, compared to −3.92, −2.90, and −2.57 for posts in the larger quartiles. The large amount of relative variation in this quartile suggests that notes attached to low-visibility posts tend to have a less consistent effect than notes on more visible posts.

\section{Factors Associated with Note Effectiveness on the Number of Views, Replies, and Likes}
\begin{figure}[H]
  \centering
  \includegraphics[width=\linewidth]{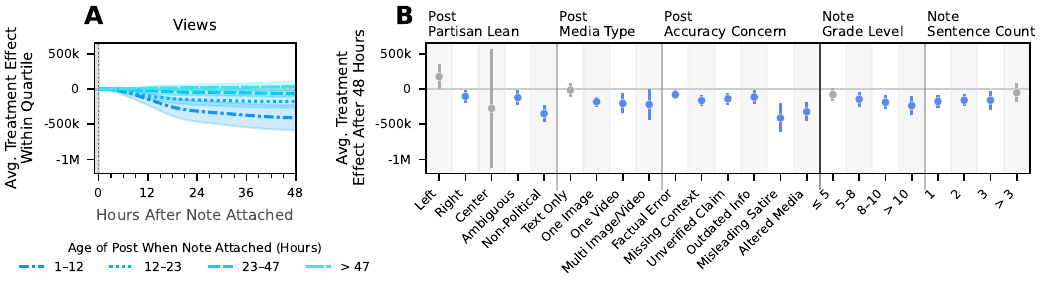}
  
  \caption{\textbf{Factors associated with effects on \textit{views}.} \textit{(A)} Average treatment effects for treated posts, stratified by hours from post creation to first note attachment. Bins represent note speed quartiles. \textit{(B)} Average treatment effects for noted posts after 48 hours, based on (i) the post's partisan lean, (ii) the number of images and videos, (iii) the accuracy concerns raised by the community note writer, (iv) the readability (Flesch–Kincaid grade level), and (v) the length of the community note. The error bands and error bars represent 95\% confidence intervals.}
  \label{fig:s1}
\end{figure}

\begin{figure}[H]
  \centering
  \includegraphics[width=\linewidth]{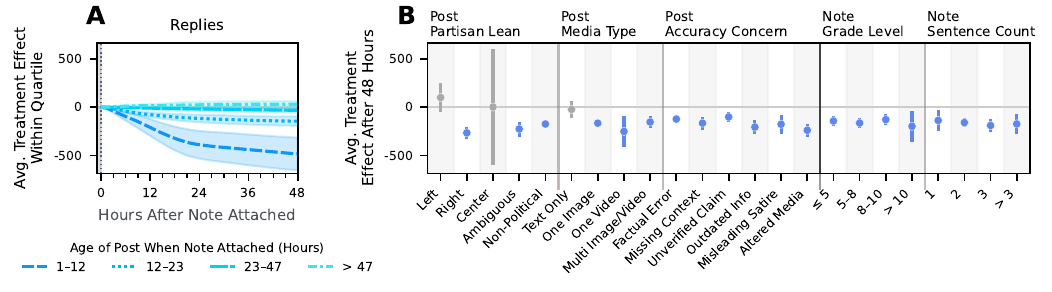}
  \scriptsize
  \caption{\textbf{Factors associated with effects on \textit{replies}}. \textit{(A)} Average treatment effects for treated posts, stratified by hours from post creation to first note attachment. Bins represent note speed quartiles. \textit{(B)} Average treatment effects for noted posts after 48 hours, based on (i) the post's partisan lean, (ii) the number of images and videos, (iii) the accuracy concerns raised by the community note writer, (iv) the readability (Flesch–Kincaid grade level), and (v) the length of the community note. The error bands and error bars represent 95\% confidence intervals.}  
  \label{fig:s2}
\end{figure}

\begin{figure}[H]
  \centering
  \includegraphics[width=\linewidth]{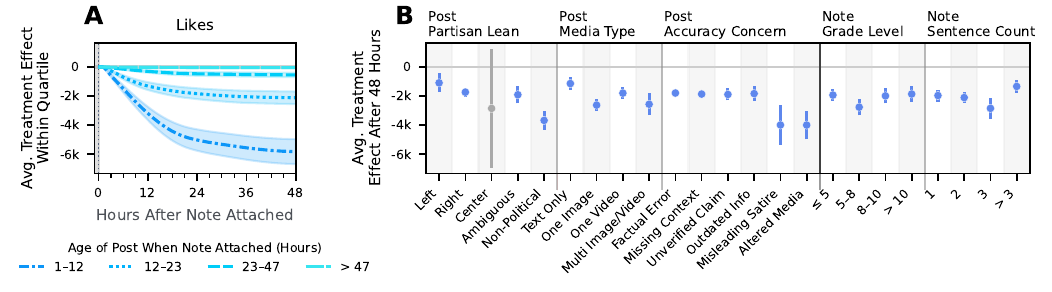}
  \scriptsize
  \caption{\textbf{Factors associated with effects on \textit{likes}}. \textit{(A)} Average treatment effects for treated posts, stratified by hours from post creation to first note attachment. Bins represent note speed quartiles. \textit{(B)} Average treatment effects for noted posts after 48 hours, based on (i) the post's partisan lean, (ii) the number of images and videos, (iii) the accuracy concerns raised by the community note writer, (iv) the readability (Flesch–Kincaid grade level), and (v) the length of the community note. The error bands and error bars represent 95\% confidence intervals.}  
  \label{fig:s3}
\end{figure}

In addition to investigating factors associated with large effects on reposts, we also examine factors associated with large effects on views, replies, and likes. We plot conditional treatment effects on views, replies, and likes in \hyperref[fig:s1]{\textit{Fig.~S3}}, \hyperref[fig:s2]{\textit{Fig.~S4}}, and \hyperref[fig:s3]{\textit{Fig.~S5}}, respectively. Most factors associated with large effects on reposts tend to show similar associations with these additional metrics; however, we note some nuances below. First, considering the partisanship of the post, we find that the treatment effects on views, replies, and likes are all larger in magnitude for right-leaning posts than for left-leaning posts. This gap appears largest for replies and views, metrics that are not as clear a positive signal of agreement from a user as a reply or a like. In fact, we estimate positive (although not statistically significant) average treatment effects on views and replies for left-leaning posts. Another trend we observe is that while the treatment effect on reposts declines as notes become more readable, the association with views is in the opposite direction. The more readable a note is, the larger the effect it tends to have on a post's reposts, but the smaller the effect it tends to have on views. This observation supports the hypothesis that unclear notes may act as a warning signal, causing users to quickly move past a post. In contrast, more comprehensible notes may engage users, making them less likely to later express agreement with the post.

\vspace{-3mm}

\vspace{-3mm}

\clearpage

\begin{figure}[t]
    \centering
    \includegraphics[width=\linewidth]{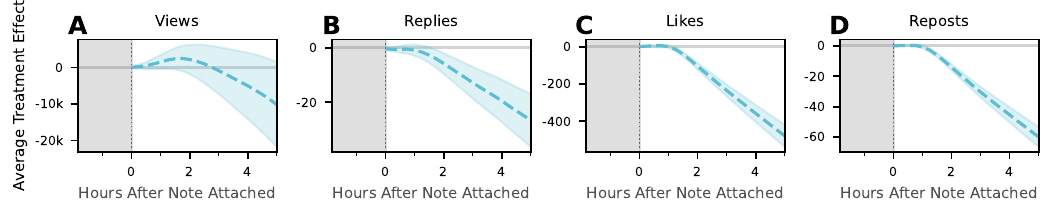}
    \caption{\textbf{Effects of note attachment on views, replies, likes, and reposts during the first five hours after a post's first note is classified as ``helpful.''}}
    \label{fig:s6}
\end{figure}

\begin{figure}[t]
    \centering
    \includegraphics[width=\linewidth]{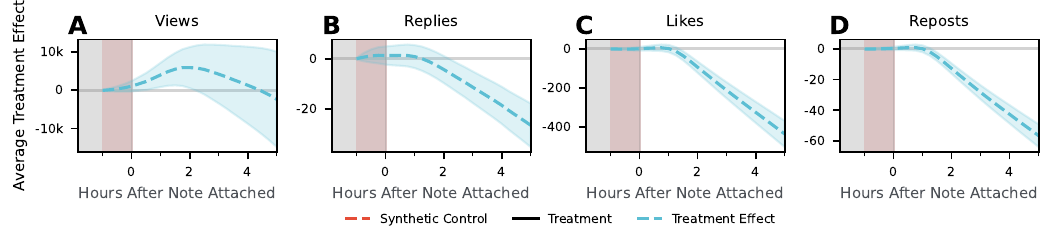}
    \caption{\textbf{One-hour placebo test: Shows the effects of note attachment on views, replies, likes, and reposts when using a one-hour backdate}. The estimated effects at the end of the backdate period---highlighted in light red, when we would expect a null effect---are 1,028 views (95\% CI: [−341, 2,397]), 1.28 replies (95\% CI: [−2.36, 4.92]), −0.154 likes (95\% CI: [−9.57, 9.26]), and 0.287 reposts (95\% CI: [−0.839, 1.41]). The slight initial increase and subsequent decrease in views is likely due to posts appearing in the ``Rated Helpful'' tab on the Community Notes webpage, which displays the most recently noted posts and is frequently visited by Community Notes volunteers.}
    \label{fig:s7}
\end{figure}

\section{In-Time Placebo Test}
\label{si:placebo-in-time}
Traditional methods for performing statistical inference are difficult to adapt to Synthetic Control Methods, and as such, researchers typically validate synthetic control-based estimates using falsification experiments~\cite{abadie_comparative_2015}. Similar to A/A tests, falsification experiments involve altering the data such that no treatment effect should be observed, before rerunning the synthetic control construction to ensure that no effect is, in fact, estimated. One such exercise is the in-time placebo test~\cite{abadie_using_2021}, first introduced by Heckman and Hotz~\cite{Heckman01121989}. This test involves ``backdating'' the treatment time, i.e., artificially considering it to have occurred earlier than it did, and then constructing the synthetic controls using only observations up until this backdated time. Any non-null treatment effects observed during the backdate period (i.e., between the backdate and the true treatment time) would suggest that the observed effects are not due to the treatment itself and would therefore undermine the credibility of the synthetic control-based treatment effect estimates. 

Prior to October 23rd, 2023, there was a delay between when a note was classified as ``helpful'' and when it was displayed on the X platform~\cite{X2023note}. This delay arises for two reasons: (1) the timestamps indicating when a note achieved a ``helpful status'' in the publicly available data---which we use to determine the treatment time in our analysis---correspond to the start time of the Community Notes algorithm’s computation, as documented in the open-source code, and (2) the time required for the algorithm’s output to propagate through X’s system and for the notes to appear on the platform. This delay typically lasts around one hour; however, there is unknown variation in its length. Therefore, we decided to consider the time when a post first received a ``helpful'' note recorded in the public Community Notes data as its treatment time. (For ease of exposition, we use the term ``note attachment'' to refer to the treatment time throughout our work, and differentiate between note attachment and treatment time only in this section.) Since the actual note attachment typically occurred about an hour after the treatment time used, our synthetic controls-based estimation of the treatment effects includes a natural in-time placebo test. 

While not easily visible in \textit{Main Text, Fig.~2} due to the axes' scales, we do find a null effect extending until approximately one hour after the first note was classified as ``helpful.'' \hyperref[fig:s6]{\textit{Fig.~S6}} shows a subset of \textit{Main Text, Fig.~2}, focusing on only the five hours following the first ``helpful'' note, rather than the full 48-hour period. As seen in the figure, we find that replies, likes, and reposts do not begin to decline meaningfully until about one hour after treatment, coinciding with the typical time when notes began to be displayed to users. The slight initial increase and subsequent decrease in views is likely due to posts appearing in the ``Rated Helpful'' tab on the Community Notes webpage, which displays the most recently noted posts and is frequently visited by Community Notes volunteers.

In addition to the natural in-time placebo test, we also perform an in-time placebo test with a one-hour backdate, estimating the synthetic controls using only observations up until one hour before a post received its first ``helpful'' note. As shown in \hyperref[fig:s7]{\textit{Fig.~S7}}, we again find that the treatment effect does not begin to decline until approximately one hour after treatment occurs. We also find that the confidence intervals estimated after backdating include zero for all metrics. One hour after the synthetic control estimation period ends (i.e., the time we consider the start of the treatment), the estimated effects were 1,028 views (95\%~CI:~[−341,~2,397]), 1.28 replies (95\% CI: [−2.36, 4.92]), −0.154 likes (95\% CI: [−9.57, 9.26]), and 0.287 reposts (95\%~CI:~[−0.839,~1.41]).

\section{Permutation Test}

\begin{figure}[t]
  \centering
  \includegraphics[width=\linewidth]{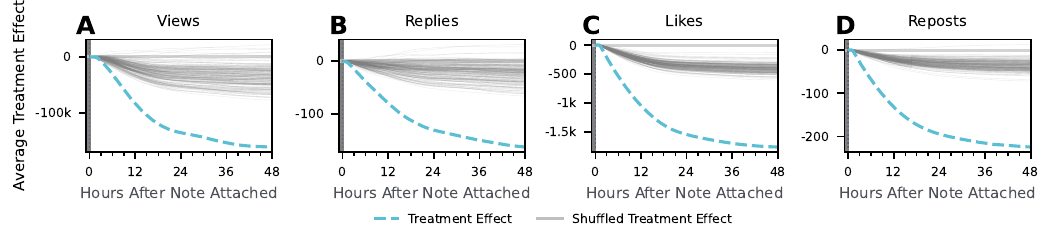}
  \caption{\textbf{Permutation test}. Each gray line shows a “placebo” effect estimated from shuffling the treatment assignment (shuffling which set of posts, from both treatment and control, are considered to be in treatment) and then measuring an average treatment effect on this artificial treatment group. To avoid cluttering the plot, we show a random sample of 200 permutations. The dashed blue lines show the average treatment effect among the posts that actually received treatment.}
  \label{fig:s8}
\end{figure}
Permutation tests are another common test of validity used with Synthetic Control Methods \cite{abadie_using_2021, abadie_penalized_2021}. Rather than testing whether a null effect is observed when artificially shifting the treatment time, as with in-time placebo tests, permutation tests artificially shuffle the treatment assignment. The complete set of units (both treatment and control) is pooled, and a random selection of the pooled units (equal in size to the original treatment group) is considered to be treated for the purposes of the artificial experiment. An average ``placebo'' effect is calculated for the artificially treated posts using all other posts as controls, regardless of whether they actually received treatment. This procedure is applied repeatedly with different samples of artificially treated posts, and a permutation distribution of average ``placebo'' effects estimated from these repetitions is compared to the observed average treatment effect. If the magnitude of the observed average treatment effect is extreme relative to the permutation distribution, it is deemed significant. A $p$-value is calculated by comparing the observed treatment effect to the average ``placebo'' effects found when shuffling the treatment assignment.

A typical permutation test accounts for two types of variability in the average treatment effect: variability due to which units receive treatment and variability due to which units are used as controls. Each unit considered to be treated is assumed to have a true, although unobserved, treatment effect. 
Treatment effects may differ across units, and a typical permutation test measures how much the average treatment effect changes when different units' treatment effects are included. This aspect of permutation testing follows logic similar to that of Fisher's exact test, comparing the observed outcome to the distribution of outcomes under random assignment. In addition to variability in the treated units, a permutation test also typically quantifies variability in the units used as controls. While each unit considered to be treated is assumed to have a true treatment effect, the synthetic control estimation process may not be able to estimate this effect precisely. By using different sets of control units in different iterations of a permutation test, one can quantify the extent to which variability in the estimation process leads to different average treatment effects on the treated units.

Permutation testing typically involves carrying out the complete treatment effect estimation process multiple times, including finding a synthetic control for each post considered to be treated in each permutation. Due to the size of our dataset, estimating average treatment effects for a single set of posts requires around one week, even with extensive parallelization across many processes, making a test involving even hundreds of iterations prohibitively long. As we are unable to perform a typical permutation test, we instead rely on the following procedure. For each unit in the pooled set, regardless of its true treatment status, we construct a synthetic control using all other units in the pooled set, including both treated and control units, as donors. The remainder of the effect estimation procedure is carried out as usual. Once treatment effects have been estimated for each unit, we then sample a set of units from the pooled set (of size equal to the number of truly treated units) and calculate an average ``placebo'' effect using these units. We perform multiple iterations of this sampling and compare the observed average treatment effect for the truly treated units to the distribution of average ``placebo'' effects estimated in these iterations. This computationally feasible procedure accounts for variability due to the units that receive treatment but does not account for variability due to the set of units used as controls.

\hyperref[fig:s8]{\textit{Fig.~S8}} shows the distribution of average ``placebo'' effects obtained by permuting the treatment assignment vector and the observed average treatment effects for the truly treated units. The average ``placebo'' effects tend to be negative because, in each permutation, some of the truly treated units---which tend to have negative individual treatment effects---are reassigned as treated. The observed average treatment effects among the truly treated units (dashed blue lines) are smaller than those reported in the main text since, as described above, the treatment effects are estimated using both treatment and control units as donors, rather than just the control units. We perform 100,000 permutations, and for each metric, we find that the observed treatment effects are strictly larger in magnitude than all average ``placebo'' effects, resulting in $p$-values of 0.00001.

\begin{figure}[t]
  \centering
  \includegraphics[width=\linewidth]{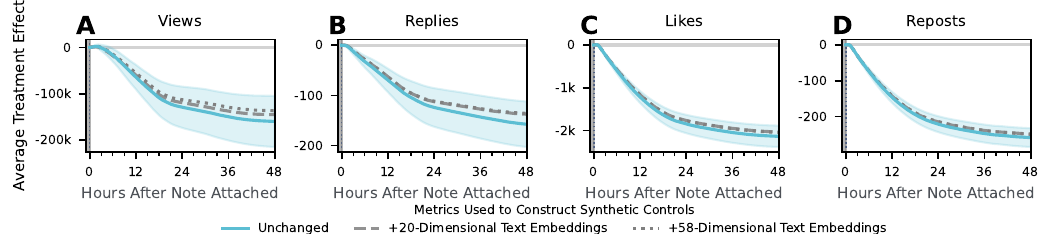}
  \caption{\textbf{Synthetic controls with additional post content embeddings}. The grey lines show the estimated treatment effects when including 20- and 58-dimensional post text embeddings in the construction of the synthetic controls, in addition to the posts' pre-treatment engagement and diffusion trajectories. All estimates fall within the 95\% confidence intervals of the original estimates based solely on the engagement and diffusion trajectories.}
  \label{fig:s10}
\end{figure}

\section{Synthetic Controls with Additional Post Content Embeddings}
In the analyses reported in the main text, we constructed a synthetic control that closely resembles each treatment post’s engagement (number of views, replies, likes, and reposts) and diffusion (repost cascade depth, breadth, and structural virality) trajectories prior to note attachment. While we restrict the donor pool for constructing the synthetic controls to posts suspected to be misleading and for which community notes have been proposed, one possible concern is that the posts which are given the largest weights in constructing the synthetic controls may be content-wise very different from the treatment post, e.g., they might be on more polarizing or controversial topics than the treatment post, which could make it harder for them to achieve the consensus required by the Community Notes algorithm for a note to be attached. To examine this concern, we perform robustness analyses that also take into account the semantic embeddings of the posts’ content in addition to their diffusion and engagement trajectories.

To generate semantic embeddings of the posts, we use Google's Gemini \texttt{text-embedding−004} model with the task type set to ``clustering,'' which produces 768-dimensional vectors per post. Before embedding the posts, we preprocess the text by removing any URLs, as most are shortened URLs that do not carry any semantic information. As the synthetic controls method involves minimizing the squared Euclidean distance between the synthetic control and treatment units, across all metrics considered, we use Principal Component Analysis (PCA) to reduce the dimensionality of the text embeddings, preventing them from dominating the distance relative to the engagement and diffusion metrics. We reduce the embeddings to two sizes: (i) 58 dimensions, considered optimal according to the method by Satopaa et al.~\cite{satopaa2011finding} for locating a ``knee'' in the curve of proportion of variance explained by each component, and (ii) 20 dimensions, a slightly smaller size included for comparison. In our main analysis, before running the synthetic controls procedure, we scale each covariate by its standard deviation among the treated units; however, in this case, we standardize the PCA-reduced text embeddings by dividing each dimension by the mean standard deviation of all dimensions for the treated units. This standardization preserves the relative importance of each PCA component---as measured by the variance explained---while giving the embeddings, as a group, the same per-dimension weight as other covariates. Manually spot-checking the outputs of the synthetic controls procedure, we observe that, in most cases, the donor posts given the largest weights are about the same claim or the same type of misinformation (e.g., vaccine-related) as the treated post, suggesting that the inclusion of the text embedding has the desired effect.

When incorporating the post text embeddings with 20 and 58 dimensions into the construction of the synthetic controls, we estimate treatment effects on views 48 hours after note attachment of −145,097 and −136,535, respectively (\hyperref[fig:s10]{\textit{Fig.~S9}}). These estimates fall well within the 95\% confidence interval for the average treatment effect on views when text embeddings are not included in the synthetic control construction, which was −159,592 (95\% CI: [−214,839, −104,344]). Similarly, the effects on replies are −138 and −136, relative to the originally estimated effect of −158  (95\% CI: [−203, −112]). For likes, the estimated treatment effects were −2,033 and −2,043, relative to the originally estimated effect of −2,134 (95\%~CI:~[−2,385,~−1,884]). Finally, for reposts, the estimated effects are −248 and −249, relative to the originally estimated effect of −259 (95\%~CI:~[−285,~−232]). These results suggest that the engagement and diffusion trajectories already capture the key information about the posts, and that incorporating post content in the construction of the synthetic controls does not significantly change the treatment effect estimates.

\section{Synthetic Controls With Donor Pools Restricted to Control Posts with High-Scoring Notes}
In the previous section, we expanded the set of covariates used to construct the synthetic controls by incorporating semantic embeddings of the posts' text, in addition to the pre-treatment engagement and diffusion trajectories. This addition allowed us to test whether ensuring that the synthetic controls are composed of control posts with similar content to the treatment post affects our treatment effect estimates. However, despite including a broad set of covariates when constructing the synthetic controls, observational causal inference inherently carries the risk that the control units may differ systematically from the treated units in unmeasured ways. As a result, the observed differences in outcomes between the two groups may reflect these unobserved confounders rather than the effect of the treatment itself.

In particular, the procedure by which community notes are selected to be shown on X creates one noteworthy way in which the treatment and control groups differ. A community note is shown on the platform only once it has been rated ``helpful'' by many users with diverse views and achieves a helpfulness score above 0.4. Therefore, it is plausible that the control posts, which never received ``helpful'' notes, do not contain the same distribution of content types as the treatment posts or otherwise differ in notable ways. As a result, they might not be a suitable pool from which to construct synthetic controls.

To test whether this discrepancy might affect our results, we limit the donor pool used to construct synthetic controls to only those control posts that received a note whose helpfulness score exceeded 0.25, 0.3, and 0.35. By placing these restrictions, we seek to limit the donor pool to control posts most similar to those in treatment, in terms of the helpfulness of their notes. Since the helpfulness scores can change over time and notes may be deleted, we rerun the Community Notes algorithm using the publicly released data from X for each day of our study period and take the maximum score each note ever achieved. Applying these score thresholds reduces the size of the donor pool from 33,321 units to 8,760 units, 5,084 units, and 2,394 units, respectively.

\begin{figure}[t]
  \centering
  \includegraphics[width=\linewidth]{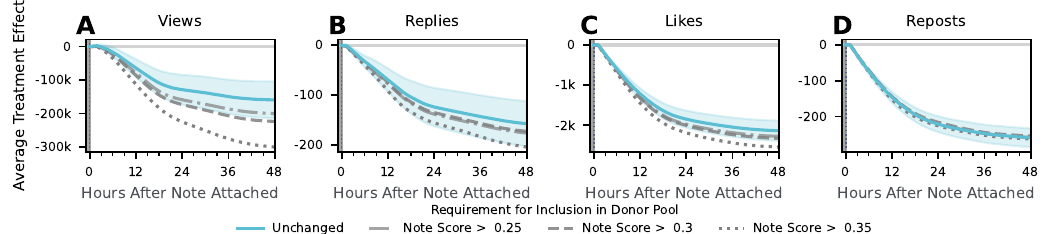}
  \caption{\textbf{Synthetic controls with donor pools restricted to control posts with high-scoring notes}. The gray lines show the estimated treatment effects when limiting the donor pool to posts with notes that have helpfulness scores above 0.25, 0.3, and 0.35. The solid blue lines and the corresponding error bands show the original point estimates and 95\% confidence intervals based on an unrestricted donor pool, as reported in the main text.}
  \label{fig:s9}
\end{figure}

When restricting the donor pool to units that receive a helpfulness score above 0.25, 0.3, and 0.35, we estimate treatment effects on views 48 hours after note attachment of −200,327, −224,108, and −300,596, respectively (\hyperref[fig:s9]{\textit{Fig.~S10}}). The estimated treatment effect when using the complete donor pool is −159,592 (95\% CI: [−214,839, −104,344]). The effects on replies are −176, −173, and −204, relative to the originally estimated effect of −158 (95\% CI: [−203, −112]). For likes, the estimated treatment effects are −2,275, −2,325, and −2,541, relative to the originally estimated effect of −2,134 (95\%~CI:~[−2,385,~−1,884]). Finally, for reposts, the estimated effects are −258, −254, and −262, relative to the originally estimated effect of −259 (95\% CI: [−285, −232]).

The larger treatment effects on views found when restricting the donor pool suggest that our original results may be overestimates rather than underestimates. However, we give a note of caution toward this interpretation. We find that using a restricted donor pool leads to larger differences between the treatment units and their synthetic controls in terms of the pre-treatment engagement and diffusion metrics. This discrepancy indicates that the smaller donor pool results in lower-quality matches, which may lead to biased estimates. Thus, we focus our presentation on the results using the full donor pool, as presented in the main text, and consider the analysis based on restricted donor pools as a robustness check.

\end{document}